\documentclass[a4paper,12pt]{article}
\usepackage{graphics}

\begin{document}

\title{Opportunity to study the LPM effect in oriented crystal
at GeV energy}
\author{V. N. Baier
and V. M. Katkov\\
Budker Institute of Nuclear Physics,\\ Novosibirsk, 630090, Russia}

\maketitle

\begin{abstract}
The spectral distribution of electron-positron pair created by
photon and the spectral distribution of photons radiated from
high-energy electron in an oriented single crystal is calculated
using the method which permits inseparable consideration both of the
coherent and incoherent mechanisms of two relevant processes. The
method includes the action of field of axis (or plane) as well as
the multiple scattering of radiating electron or particles of the
created pair (the Landau-Pomeranchuk-Migdal (LPM) effect). The
influence of scattering on the coherent mechanism and the influence
of field on the incoherent mechanism are analyzed. In tungsten, axis
$<111>$ for the pair creation process at temperature T= 100 K the
LPM effect attains 8 \% at photon energy 5 GeV and for the radiation
process at T= 293 K the LPM effect reaches 6 \% at electron energy
10 GeV.

\end{abstract}

\newpage

\section{Introduction}

When the radiation formation length becomes comparable to the
distance over which the multiple scattering becomes important, the
probability of the bremsstrahlung process will be suppressed. This
is is the Landau- Pomeranchuk -Migdal (LPM) effect. The
characteristic energy $\varepsilon_e$ when the LPM effect affects
the whole spectrum is
\begin{eqnarray}
&&\varepsilon_e=\frac{m}{16\pi Z^2\alpha^2\lambda_c^3n_aL_0}, \quad
L_0=\ln(ma)+ \frac{1}{2}-f(Z\alpha),\quad a=\frac{111Z^{-1/3}}{m},
\nonumber \\
&& \quad f(\xi)=\sum_{n=1}^{\infty}
\frac{\xi^2}{n(n^2+\xi^2)},\label{1}
\end{eqnarray}
where $Z$ is the charge of nucleus, $n_a$ is the mean atom density,
$f(\xi)$ is the Coulomb correction. The energy $\varepsilon_e$ is
very high even for heavy elements: $\varepsilon_e=2.73$~TeV  for
tungsten and $\varepsilon_e=4.38$~TeV for lead.

The LPM effect was studied in SLAC E-146 experiment in many elements
using electrons with energy 8~GeV and 25~GeV \cite{ABB} and in
CERN-SPS experiment in iridium Ir ($\varepsilon_e=2.27$~TeV)  and
tantalum Ta ($\varepsilon_e=3.18$~TeV) for electrons with energy up
to 287 GeV \cite{HU}. The suppression of bremsstrahlung was observed
in the soft part of spectrum $\omega \ll \varepsilon$. The LPM
effect will manifest itself also in the process of electron-positron
pair creation by a photon. In this process the characteristic photon
energy $\omega_e=4\varepsilon_e$. In contrast to the radiation
process, where the LPM effect can be observed in soft part of
spectrum, for observation of the effect in the pair creation process
the energy $\omega \sim \omega_e$ is needed. No such energy is
available for the time being. For description of the LPM effect
including the discussion of mentioned experiments see e.g. review
\cite{BK}.

Recently authors developed a new approach to analysis of pair
creation by a photon \cite{BK0} and radiation from high energy
electrons \cite{BK1} in oriented crystals . This approach not only
permits to consider simultaneously both the coherent and incoherent
mechanisms of pair creation by a photon or photon emission from high
energy electrons but also gives insight on influence of the LPM
effect on the considered mechanisms of pair creation or radiation in
oriented crystals. It is shown that the relative contribution of the
LPM effect in total probability of pair creation $\Delta_p$
\begin{equation}
\Delta_p=-\frac{W - W^{coh} -W^{inc}}{W}, \label{1a}
\end{equation}
where $W$ is the total probability of pair creation, $W^{coh}$ is
the total coherent probability of pair creation and  $W^{inc}$ is
the total incoherent probability of pair creation. In tungsten
crystal (axis $<111>$) the relative contribution of the LPM effect
attains $\Delta_p \simeq 5.5$ \% at the photon energy $\omega \simeq
7$~GeV for the temperature T=100 K and $\Delta_p \simeq 4.3$\% at
$\omega \simeq 12$~GeV for T=293 K \cite{BK0}. The origin of the
effect is connected with high effective density of atomic nuclei
near the crystalline axis which can exceed the  mean atom density by
3 order of magnitude. At higher photon energy the field action
excludes the LPM effect. As opposed to the pair creation process
where the coherent process is suppressed exponentially at the low
photon energy, the coherent contribution to the radiation intensity
is very essential for any electron energy. Because of this the
relative contribution of the LPM effect into the total radiation
intensity (the inverse radiation length) $\Delta_r$ ($\Delta_r$ is
obtained from $\Delta_p$ by substitution of the probability $W$ in
Eq.(\ref{1a}) by the corresponding radiation intensity) is much
smaller than in the pair creation probability: in tungsten crystal
(axis $<111>$) it has the maximum $\Delta_r \simeq 0.9$\% at the
electron energy $\varepsilon \simeq 0.3$~GeV for T=100 K and
$\Delta_r \simeq 0.8$\% at $\varepsilon \simeq 0.7$~GeV for T=293 K
\cite{BK1}.

As it was mentioned, the LPM effect in an amorphous medium was
observed in the radiation spectra. It turns out that in an oriented
crystal the study of the pair creation  spectrum (or the hard part
of the radiation spectrum) is very convenient approach for the
investigation of the LPM effect.

The study performed in the present paper is connected with
experiment NA63 carried out recently at SPS at CERN  (for proposal
see \cite{NA63}).

\section{Spectrum of particles of a pair created by a photon}

Basing on Eqs.(16) and (17) of \cite{BK0} (see also Eq.(7.135) in
\cite{BKS}) one get the general expression for the spectral
distribution of particles of pair created by a photon
\begin{eqnarray}
&& dW(\omega, y)=\frac{\alpha m^2}{2\pi \omega} \frac{dy}{y(1-y)}
\int_0^{x_0}\frac{dx}{x_0}G(x, y),\quad G(x, y)=\int_0^{\infty} F(x,
y, t)dt +s_3\frac{\pi}{4},
\nonumber \\
&& F(x, y, t)={\rm Im}\left\lbrace e^{f_1(t)}\left[s_2\nu_0^2
(1+ib)f_2(t)-s_3f_3(t) \right] \right\rbrace,\quad
b=\frac{4\kappa_1^2}{\nu_0^2}, \quad y=\frac{\varepsilon}{\omega},
\nonumber \\
&& f_1(t)=(i-1)t+b(1+i)(f_2(t)-t),\quad
f_2(t)=\frac{\sqrt{2}}{\nu_0}\tanh\frac{\nu_0t}{\sqrt{2}},
\nonumber \\
&&f_3(t)=\frac{\sqrt{2}\nu_0}{\sinh(\sqrt{2}\nu_0t)}, \label{2}
\end{eqnarray}
where
\begin{equation}
s_2=y^2+(1-y)^2,~s_3=2y(1-y),~\nu_0^2=4y(1-y)
\frac{\omega}{\omega_c(x)},~\kappa_1=y(1-y)\kappa(x), \label{3}
\end{equation}
$\varepsilon$ is the energy of one of the created particles.

The situation is considered when the photon angle of incidence
$\vartheta_0$ (the angle between photon momentum {\bf k} and the
axis (or plane)) is small $\vartheta_0 \ll V_0/m$. The axis
potential (see Eq.(9.13) in \cite{BKS}) is taken in the form
\begin{equation}
U(x)=V_0\left[\ln\left(1+\frac{1}{x+\eta} \right)-
\ln\left(1+\frac{1}{x_0+\eta} \right) \right], \label{4}
\end{equation}
where
\begin{equation}
x_0=\frac{1}{\pi d n_a a_s^2}, \quad  \eta_1=\frac{2
u_1^2}{a_s^2},\quad x=\frac{\varrho^2}{a_s^2}, \label{5}
\end{equation}
Here $\varrho$ is the distance from axis, $u_1$ is the amplitude of
thermal vibration, $d$ is the mean distance between atoms forming
the axis, $a_s$ is the effective screening radius of the potential.
The parameters in Eq.(\ref{4}) were determined by means of fitting
procedure, see Table 1.

The local value of parameter $\kappa(x)$  which determines the
probability of pair creation in the field Eq.(\ref{4}) is
\begin{equation}
\kappa(x)=-\frac{dU(\varrho)}{d\varrho}\frac{\omega}{m^3}=2\kappa_sf(x),\quad
f(x)=\frac{\sqrt{x}}{(x+\eta)(x+\eta+1)},\quad \kappa_s=\frac{V_0
\omega}{m^3a_s}\equiv \frac{\omega}{\omega_s}.
 \label{6}
\end{equation}
For an axial orientation of crystal the ratio of the atom density
$n(\varrho)$ in the vicinity of an axis to the mean atom density
$n_a$ is (see \cite{BK0})
\begin{equation}
\frac{n(x)}{n_a}=\xi(x)=\frac{x_0}{\eta_1}e^{-x/\eta_1},\quad
\omega_0=\frac{\omega_e}{\xi(0)}, \quad
\omega_e=4\varepsilon_e=\frac{m}{4\pi
Z^2\alpha^2\lambda_c^3n_aL_0}.\label{7}
\end{equation}

The functions and values in Eqs.(\ref{2}) and (\ref{3}) are
\begin{eqnarray}
&&\omega_c(x)=
\frac{\omega_e(n_a)}{\xi(x)g_p(x)}=\frac{\omega_0}{g_p(x)}e^{x/\eta_1},\quad
L=L_0g_p(x),
\nonumber \\
&&
 g_p(x)=g_{p0}+\frac{1}{6 L_0}\left[\ln
\left(1+\kappa_1^2\right)+\frac{6 D_{p}\kappa_1^2}
{12+\kappa_1^2}\right],\quad
g_{p0}=1-\frac{1}{L_0}\left[\frac{1}{42}+h\left(\frac{u_1^2}{a^2}\right)\right],
\nonumber \\
&&  h(z)=-\frac{1}{2}\left[1+(1+z)e^{z}{\rm Ei}(-z)
\right],\label{8}
\end{eqnarray}
where $L_0$ is defined in Eq.(\ref{1}), the function $g_p(x)$
determines the effective logarithm using the interpolation
procedure, $D_{p}=D_{sc}-10/21=1.8246$, $D_{sc}=2.3008$ is the
constant entering in the radiation spectrum at $\chi/u \gg 1$ (or in
electron spectrum in pair creation process at $\kappa_1 \gg 1$), see
Eq.(7.107) in \cite{BKS},~ Ei($z$) is the integral exponential
function.

The expression for $dW(\omega, y)$ Eq.(\ref{2}) includes both the
coherent and incoherent contributions as well as the influence of
the multiple scattering (the LPM effect) on the pair creation
process (see \cite{BK0}). The probability of the coherent pair
creation is the first term ($\nu_0^2=0$) of the decomposition of
Eq.(\ref{2}) over $\nu_0^2$ (compare with Eq.(12.7) in \cite{BKS})
\begin{eqnarray}
&& dW^{coh}(\omega, y)=\frac{\alpha
m^2}{2\sqrt{3}\pi\omega}\frac{dy}{y(1-y)}
\int_0^{x_0}\frac{dx}{x_0}\left[2s_2K_{2/3}(\lambda)
+s_3\int_{\lambda}^{\infty}K_{1/3}(z)dz \right],
\nonumber \\
&&\lambda=\lambda(x)=\frac{2}{3\kappa_1}, \label{9}
\end{eqnarray}
where $K_{\nu}(\lambda)$ is MacDonald's function. The probability of
the incoherent pair creation is the second term ($\propto \nu_0^2$)
of the mentioned decomposition (compare with Eq.(21.31) in
\cite{BKS})
\begin{equation}
dW^{inc}(\omega, y)=\frac{4Z^2\alpha^3 n_a L_0}{15 m^2} dy
\int_0^{\infty}\frac{dx}{\eta_1}e^{-x/\eta_1}f(x, y)g_p(x),
\label{10}
\end{equation}
where $g_p(x)$ is defined in Eq.(\ref{8}),
\begin{eqnarray}
&&f(x, y)=f_1(z)+ s_2f_2(z), \quad
f_1(z)=z^4\Upsilon(z)-3z^2\Upsilon'(z)-z^3,
\nonumber \\
&& f_2(z)=(z^4+3z)\Upsilon(z)-5z^2\Upsilon'(z)-z^3,\quad z=z(x,
y)=\kappa_1^{-2/3}. \label{11}
\end{eqnarray}
Here
\begin{equation}
\Upsilon(z)=\int_0^{\infty}\sin\left(zt+\frac{t^3}{3}\right)dt
\label{12}
\end{equation}
is the Hardy function.

The next terms of decomposition of the pair creation probability
$dW=dW(\omega, y)$ over $\nu_0^2$ describe  the influence of
multiple scattering on the pair creation process, the LPM effect.
The third term ($\propto \nu_0^4$) of the mentioned decomposition
has the form
\begin{eqnarray}
&& \frac{dW^{(3)}(\omega, y)}{dy}=-\frac{\alpha
m^2\omega\sqrt{3}}{5600 \pi \omega_0^2 x_0}
\int_0^{x_0}\frac{g_p^2(x)}{\kappa(x)}\Phi(\lambda)e^{-2x/\eta_1}dx
\nonumber \\
&&
\Phi(\lambda)=\lambda^2\left(s_2F_2(\lambda)-s_3F_3(\lambda)\right),
\nonumber \\
&& F_2(\lambda)=(7820+126\lambda^2)\lambda K_{2/3}(\lambda)-
(280+2430\lambda^2) K_{1/3}(\lambda),
\nonumber \\
&& F_3(\lambda)=(264-63\lambda^2)\lambda K_{2/3}(\lambda)-
(24+3\lambda^2) K_{1/3}(\lambda),\label{13}
\end{eqnarray}
where $\lambda$ is defined in Eq.(\ref{9}).

We consider the case of relatively low photon energies where the
influence of the field of axis on the pair creation process is still
weak. In this case it is possible to single out the basic elements
which distinguish the pair creation process in oriented crystal from
the same process in amorphous medium. The first of these elements is
the modification of the characteristic logarithm $L_0$ Eq.(\ref{1})
\begin{equation}
\ln a \rightarrow \ln a - h\left(\frac{u_1^2}{a^2}\right).
\label{14}
\end{equation}
For $u_1\ll a$ one has $h(u_1^2/a^2) \simeq
-(1+C)/2+\ln(a/u_1),~C=0.577..$ and so this term characterizes the
new value of upper boundary of impact parameters $u_1$ contributing
to the value $<{\bf q}_s^2>$ instead of screening radius $a$ in an
amorphous medium (${\bf q}_s$ is the momentum transfer at random
collision, see \cite{BK0}).

Because to action of the field of axis the coherent pair creation by
a photon process emerges at $\omega \sim \omega_m$. In
near-threshold region the probability of this process has the form
(see Eq.(12.14) in \cite{BKS})
\begin{eqnarray}
&& dW^{coh}=\frac{\alpha m^2 dy}{\omega_m
x_0}\sqrt{-\frac{3f(x_m)}{4f''(x_m)}}\left(1-\frac{s}{4}\right)
\exp\left(-\frac{8}{3\kappa_m s}\right)
\nonumber \\
&& \kappa_m=\kappa(x_m)\equiv \frac{\omega}{\omega_m}, \quad
\kappa'(x_m)=0,\quad
x_m=\frac{1}{6}(\sqrt{1+16\eta(1+\eta)}-1-2\eta)
\nonumber \\
&& s=s(y)=4y(1-y),\label{15}
\end{eqnarray}
where the functions $f(x), \kappa(x)$ are defined in Eq.(\ref{6}).

The correction to the probability of the incoherent pair creation in
the region of the weak influence of the axis field is positive and
the probability itself is
\begin{eqnarray}
&&dW^{inc}=dW^{cr}\left[1+\frac{5}{8}\sigma_1s^2\left(1+
\frac{7s}{150d}\right)\right];\quad dW^{cr}=\frac{\alpha m^2 \eta_1
d }{\pi \omega_g x_0}dy,\quad \omega_g=\frac{\omega_0}{g_{p0}},
\nonumber \\
&& d=d(y)=1-\frac{s(y)}{3},\quad
\sigma_n=\sigma_n(\omega)=\int_{0}^{x_0}\kappa^2(x)
\exp\left(-\frac{nx}{\eta_1}\right)\frac{dx}{\eta_1}, \label{16}
\end{eqnarray}
where $g_{p0}$ is defined in Eq.(\ref{8}). Using Eq.(\ref{7}) one
has
\begin{equation}
dW^{cr}=\frac{\alpha m^2 \eta_1}{\pi \omega_0 x_0}d(y)
g_{p0}dy=\frac{4Z^2\alpha^3}{m^2}n_aL_0d(y)g_{p0}dy \label{17}
\end{equation}
If one omits in this expression the crystal summand in
$g_{p0}~(h(u_1^2/a^2))$, the probability $dW^{cr}$ will be very
close to the Bethe-Maximon probability.

The expression Eq.(\ref{13}) for $dW^{(3)}/dy$ contains the same
near-threshold smallness as in Eq.(\ref{15}) and additionally the
small factor $\nu_0^4$. Because of this one can neglect this term in
the region of applicability of Eqs.(\ref{15}),(\ref{16}).

The next terms of decomposition of the pair creation probability
$dW=dW(\omega, y)$ over $\nu_0^2$ in the region under study is
\begin{equation}
\frac{dW^{(4)}}{dy}=-\frac{dW^{cr}}{dy}\frac{\omega^2s^2}{3\omega_g^2}
\left[\left(1+\frac{5s}{63d}\right)\left(1+\frac{377}{16}\sigma_3s^2\right)
-\frac{1651\sigma_3s^2}{10080d}\right]. \label{18}
\end{equation}
The last equation, which describes the LPM effect in the region of
weak influence of the field, has the rather narrow region of
applicability because of the large coefficient 377/16 in front of
depending on the field correction. Let us note that one can use
following simple expressions for the entering $\sigma_{1,3}$
\begin{equation}
\sigma_1 \simeq \frac{3}{4}\left(\frac{\omega}{\omega_m}\right)^2,
\quad \sigma_3 \simeq
\frac{3}{14}\left(\frac{\omega}{\omega_m}\right)^2,  \label{19}
\end{equation}
without violating the accuracy of derived above approximate
probabilities.

 The different contributions to the spectra of
created pair (in units ${\rm cm}^{-1}$) in tungsten, axis $<111>$,
temperature T=100 K, for the energies where the coherent and the
incoherent contributions are comparable, are shown in Fig.1(a),
where one-half of spectra, which are symmetric with respect of the
line $y=0.5$, are shown. Let us discuss the spectra. When one of the
created particles is soft $y \ll 1$ (the other particle takes the
large part of photon energy) the incoherent contributions dominate.
For $\kappa_m \geq 1$ and at $y \ll 1/\kappa_m$ this part of the
spectrum is described by Eqs.(\ref{15}), (\ref{16}). With $y$
increase the coherent contributions appear. Their relative
contributions to the summary spectra grow fast with photon energy
increase: if for $\omega=5~$GeV (the lowest considered energy) the
coherent contribution is rather small, then for $\omega=15~$GeV (the
highest considered energy) the coherent contribution dominates at
$y\sim 0.5$. In this region the incoherent contributions decrease.
This reduction becomes more essential with photon energy increase.
For $\omega=7~$GeV the interplay of the coherent and incoherent
contributions is leading to the nearly flat final spectrum (the
variation is less than 10 \%, this is quite unusual). It should be
noted that for $\omega=7~$GeV the right end ($y=0.5$) is slightly
lower than the left end of spectrum ($y\rightarrow 0$):
$dW/dy(y\rightarrow 0)=2.303~{\rm cm}^{-1}$ and
$dW(y=0.5)=2.215~{\rm cm}^{-1}$, while the sum of the incoherent and
coherent contributions is slightly higher:
$dW^{inc}/dy(y=0.5)+dW^{coh}/dy(y=0.5)=2.365~{\rm cm}^{-1}$. The
arising difference is the consequence of the LPM effect. This
property may be very useful in experimental study.

We define the contribution of the LPM effect into the spectral
distribution of created pair, by analogy with \cite{BK0}, as
\begin{equation}
\Delta_p(\omega, y)=-\frac{dW(\omega, y) -dW^{coh}(\omega,
y)-dW^{inc}(\omega, y)}{dW(\omega, y)}. \label{20}
\end{equation}
The function $\Delta_p(\omega, y)$ is shown in Fig.1(b). It reaches
the highest value $\Delta_p=$ 8.35 \% at $\omega=5~$GeV and $y=0.5$.
At $\omega=7~$GeV the maximal value $\Delta_p=$ 8.15 \% is attained
at $y=0.24$, at $\omega=10~$GeV the maximal value $\Delta_p=$ 8 \%
is attained at $y=0.16$, and at $\omega=15~$GeV the maximal value
$\Delta_p=$ 7.86 \% is attained at $y=0.09$.

The same characteristics but for germanium, axis $<110>$, T= 293 K
are shown in Fig.2. In this case the nearly flat spectrum appears at
$\omega=55~$GeV. The LPM effect is essentially weaker: at
$\omega=55~$GeV the maximal value $\Delta_p=$ 2.04 \% is attained at
$y=0.34$, at $\omega=75~$GeV the maximal value $\Delta_p=$ 2 \% is
attained at $y=0.22$, and at $\omega=95~$GeV the maximal value
$\Delta_p=$ 1.98 \% is attained at $y=0.16$.

All the curves in Figs.1(b) and 2(b) have nearly the same height of
the maximum and the position of the maximum $y_m$ can be found
roughly by solving the equation $s(y_m)=2\omega_m/3\omega$ (the
function $s(y)$ is defined in Eq.(\ref{15})). Since $s(y_m) \leq 1$
one has from the equation the boundary value of the photon energy
$\omega_b \simeq 2\omega_m/3$. For higher photon energy the value
$\Delta_p(\omega, y)$ varies insignificantly. At $\omega \ll
\omega_b$ the value $\Delta_p^{max}=\Delta_p(\omega, 1/2)$ decreases
as $\omega^2$ with $\omega$ reduction according to Eq.(\ref{18})).
The absolute maximum of the LPM effect is achieved at $\omega=
\omega_b$ and $y=1/2$. Since at $\omega \geq \omega_b$
Eq.(\ref{18})) is unapplicable at $y=y_m$ (in the maximum of the LPM
effect the depending on field correction is large
(($377/16)\sigma_3(\omega)s^2(y_m) \simeq 5s^2(y_m)\kappa_m^2 \simeq
2$), so Eq.(\ref{18})) can be used for rough estimates only:
$\Delta_p^{max} \sim \omega_m^2/3\omega_g^2$. For tungsten, T=100 K
one has $\omega_m/\omega_g \simeq 0.54$, so that $\Delta_p^{max}
\sim 9\%$ (in reasonable agreement with Fig.1(b)); the position
estimates according to the presented scheme are in good agreement
with Fig.1(b). For tungsten, T=293 K one has $\omega_m/\omega_g
\simeq 0.43$, so that $\Delta_p^{max} \sim 6\%$; the numerical
calculation in frame of the developed theory gives for the pair
($\Delta_p, y_m$) the following results: for $\omega=10~$GeV (6.6
\%, 0.36), for $\omega=15~$GeV (6.4 \%, 0.18), for $\omega=25~$GeV
(6.3 \%, 0.1). The position estimates according to presented scheme
are in reasonable agreement with these data. For germanium, T=293 K
one has $\omega_m/\omega_g \simeq 1/7$, so the magnitude of the LPM
is essentially smaller. Small magnitude of the LPM effect for light
and intermediate elements was discussed in \cite{BK}.

\section{Spectrum of radiation from high-energy electron}

The expression for the spectral probability of radiation used in the
above derivation can be found from the spectral distribution
Eq.(\ref{2}) ($dW/dy=\omega dW/d\varepsilon $) using the standard
QED substitution rules: $\varepsilon \rightarrow
-\varepsilon,~\omega \rightarrow -\omega,~\varepsilon^2d\varepsilon
\rightarrow \omega^2d\omega$ and exchange $\omega_c(x) \rightarrow
4\varepsilon_c(x)$. As a result one has for the spectral intensity
$dI=\omega dW$
\begin{eqnarray}
&& dI(\varepsilon,y_r)=\frac{\alpha m^2}{2\pi} \frac{y_r
dy_r}{1-y_r} \int\limits_0^{x_0}\frac{dx}{x_0}G_{r}(x, y_r),
\nonumber \\
&&G_{r}(x, y_r)=\int\limits_0^{\infty} F_{r}(x, y_r, t)dt
-r_{3}\frac{\pi}{4},
\nonumber \\
&& F_{r}(x, y_r, t)={\rm Im}\left\lbrace
e^{\varphi_1(t)}\left[r_{2}\nu_{0r}^2 (1+ib_r)f_2(t)+r_{3}f_3(t)
\right] \right\rbrace,\quad b_r=\frac{4\chi^2(x)}{u^2\nu_{0r}^2},
\nonumber \\
&& y_r=\frac{\omega}{\varepsilon}, \quad u=\frac{y_r}{1-y_r},\quad
\varphi_1(t)=(i-1)t+b_r(1+i)(f_2(t)-t), \label{r1}
\end{eqnarray}
where
\begin{eqnarray}
&&r_2=1+(1-y_r)^2,\quad r_3=2(1-y_r),\
\nonumber \\
&&\nu_{0r}^2=\frac{1-y_r}{y_r} \frac{\varepsilon}{\varepsilon_c(x)},
\label{r2}
\end{eqnarray}
where the functions $f_2(t)$ and $f_3(t)$ are defined in
Eq.(\ref{2}). The local value of parameter $\chi(x)$  which
determines the radiation probability in the field Eq.(\ref{4}) is
\begin{equation}
\chi(x)=-\frac{dU(\varrho)}{d\varrho}\frac{\varepsilon}{m^3}=2\chi_s
f(x),\quad \chi_s=\frac{V_0 \varepsilon}{m^3a_s}\equiv
\frac{\varepsilon}{\varepsilon_s},
 \label{r3}
\end{equation}
where $f(x)$ is defined in Eq.(\ref{6}).

The functions and values in Eqs.(\ref{r1}) and (\ref{r2}) (see also
Eqs.(\ref{7}) and (\ref{8})) are
\begin{eqnarray}
&&\varepsilon_c(x)=
\frac{\varepsilon_e(n_a)}{\xi(x)g_r(x)}=\frac{\varepsilon_0}{g_r(x)}e^{x/\eta_1},
\nonumber \\
&&g_r(x)=g_{r0}+\frac{1}{6 L_0}\left[\ln
\left(1+\frac{\chi^2(x)}{u^2}\right)+\frac{6 D_{r}\chi^2(x)}
{12u^2+\chi^2(x)}\right],
\nonumber \\
&&
g_{r0}=1+\frac{1}{L_0}\left[\frac{1}{18}-h\left(\frac{u_1^2}{a^2}\right)\right],\quad
\label{r4}
\end{eqnarray}
where the function $g_r(x)$ determines the effective logarithm using
the interpolation procedure:$L=L_0g_r(x)$, see Eq.(\ref{8}),
$D_r=D_{sc}-5/9$=1.7452.

The expression for $dI$ Eq.(\ref{r1}) includes both the coherent and
incoherent contributions as well as the influence of the multiple
scattering (the LPM effect) on the photon emission process (see
\cite{BK1}).  The intensity of the coherent radiation is the first
term ($\nu_0^2=0$) of the decomposition of Eq.(\ref{r1}) over
$\nu_{0r}^2$ (compare with Eq.(17.7) in \cite{BKS})
\begin{eqnarray}
&&dI^{coh}(\varepsilon,y_r)=\frac{\alpha m^2}{\sqrt{3}\pi}\frac{y_r
dy_r}{1-y_r}
\int\limits_0^{x_0}\frac{dx}{x_0}\left[r_2K_{2/3}(\lambda_r)
-(1-y_r)\int_{\lambda_r}^{\infty}K_{1/3}(z)dz \right],
\nonumber \\
&& \lambda_r=\lambda_r(x)=\frac{2u}{3\chi(x)}. \label{r5}
\end{eqnarray}
The intensity of the incoherent radiation is the second term
($\propto \nu_0^2$) of the mentioned decomposition (compare with
Eq.(21.21) in \cite{BKS})
\begin{equation}
dI^{inc}(\varepsilon,y_r)=\frac{\alpha m^2}{60 \pi}
\frac{\varepsilon}{\varepsilon_0}dy_r
\int\limits_0^{\infty}\frac{dx}{x_0}e^{-x/\eta_1}f_r(x, y_r)g_r(x),
\label{r6}
\end{equation}
where
\begin{eqnarray}
&&f_r(x, y_r)=\left[y_r^2(f_1(z)+f_2(z))+2(1-y_r)f_2(z)\right],
\nonumber \\
&&z=\left(\frac{u}{\chi(x)}\right)^{2/3}, \label{r7}
\end{eqnarray}
the functions $f_{1,2}(z)$ are defined in Eq.(\ref{11}).

The next terms of decomposition of for the spectral intensity of
radiation $I(\varepsilon,y_r)$ over $\nu_{0r}^2$ describe the
influence of multiple scattering on the photon emission process, the
LPM effect. The third term ($\propto \nu_{0r}^4$) of the mentioned
decomposition has the form
\begin{equation}
dI^{(3)}(\varepsilon, y_r)=-\frac{\alpha m^2\sqrt{3}}{89600\pi x_0}
dy_r \left(\frac{\varepsilon}{\varepsilon_0}\right)^2
\int\limits_0^{x_0}\frac{g_r^2(x)}{\chi(x)}
\Phi(\lambda_r(x))e^{-2x/\eta_1}dx,\label{r8}
\end{equation}
where
\begin{equation}
\Phi(\lambda_r)=\lambda_r^2(r_2F_2(\lambda_r)+r_3F_3(\lambda_r)),\label{r9}
\end{equation}
where $\lambda_r$ is defined in Eq.(\ref{r5}), the functions $F_2$
and $F_3$ are defined in Eq.(\ref{13}).

When $\chi_m=\chi(x_m)\equiv \varepsilon/\varepsilon_m \leq 1$ (see
Eq.(\ref{15})) and the emitted photon is soft enough ($y_r \ll
\chi_m$) Eqs.(17.11)-(17.13) in \cite{BKS} may be used. For
$\vartheta_0=0$ and $u \simeq y_r \ll 1$ one has the following
expression
\begin{equation}
\frac{dI^{coh}}{d\omega}=\left(\frac{2}{\sqrt{3}}\right)^{5/3}
\Gamma\left(\frac{2}{3}\right)\frac{\alpha m^2}{\pi x_0
\varepsilon_s} \left(\frac{y_r}{\chi_s}\right)^{1/3}\left(
\ln\left(\frac{\chi_s}{y_r}\right)+a(\eta)\right),\label{r10}
\end{equation}
where
\begin{eqnarray}
&&a=a(\eta)=\ln(18\sqrt{3})-\frac{\pi}{2\sqrt{3}}-C-\frac{3}{4}-
l_1(\eta), \quad C=0.577...
\nonumber \\
&&l_1(\eta)=3.975\beta^{2/3}\left(1+\frac{8\beta}{15}+\frac{7
\beta^2}{18}\right)-\beta\left(\frac{3}{2}+\frac{9\beta}{8}+
\frac{13\beta^2}{14}\right),~  \beta=\frac{\eta}{1+\eta}.
\label{r11}
\end{eqnarray}
The position of the maximum $y_{rm}$ of this contribution and its
value are given by the expressions
\begin{equation}
y_{rm}=\exp(-3-a(\eta))\chi_s, \quad
\frac{dI^{coh}(y_{rm})}{d\omega}\simeq
\frac{8\varepsilon_0}{\varepsilon_s \eta_1
L_{rad}}\left(1+\frac{2}{3}a\right)e^{-a/3}, \label{r12}
\end{equation}
where $L_{rad}$ is the Bethe-Maximon radiation length, see e.g.
Eq.(7.54) in \cite{BKS}).

On the same assumptions the contribution of the incoherent radiation
is (see Eq.(7.107) in \cite{BKS})
\begin{eqnarray}
&&\frac{dI^{inc}}{d\omega}=\Gamma\left(\frac{1}{3}\right)\frac{1}{5L_{rad}}
\int\limits_0^{x_0}\left(\frac{y_r}{3\chi(x)}\right)^{2/3}\left[g_{r0}
+\frac{1}{L_0}\left(D_r+\frac{1}{3}\ln\frac{\chi(x)}{y_r}\right)\right]
e^{-x/\eta_1}\frac{dx}{\eta_1}
\nonumber \\
&& \simeq 0.3 L_{rad}^{-1}\left(\frac{y_r}{\chi_m}\right)^{2/3}
\left[g_{r0}+\frac{1}{L_0}\left(\frac{5}{3}+\frac{1}{3}\ln
\frac{\chi_m}{y_r}\right)\right] \label{r13}
\end{eqnarray}
In the maximum of the spectral distribution this contribution can be
written as
\begin{equation}
\frac{dI^{inc}}{d\omega} \simeq 0.04
\left(\frac{\varepsilon_m}{\varepsilon_s}\right)^{2/3}e^{-2a/3}
\left[g_{r0}+\frac{1}{L_0}\left(\frac{8}{3}+\frac{a}{3}+\frac{1}{3}\ln
\frac{\varepsilon_s}{\varepsilon_m}\right)\right]L_{rad}^{-1},
\label{r14}
\end{equation}
and it is very small comparing with the coherent one.

In the case of weak influence of the axis field the intensity
spectrum of the incoherent radiation has the form
\begin{eqnarray}
&&\frac{dI^{inc}}{d\omega} \simeq \frac{dI^{cr}}{d\omega}
\left[1+\frac{15}{2}\left(\frac{\chi_m}{u}\right)^2
\left(1-\frac{14}{75}\frac{(1-y_r)}{d_r(y_r)}\right)\right],\quad
d_r(y_r)=y_r^2+\frac{4}{3}(1-y_r),
\nonumber \\
&& \frac{dI^{cr}}{d\omega}= \frac{\alpha m^2 \eta_1}{4\pi x_0}
\frac{\varepsilon}{\varepsilon_g}d_r(y_r),\quad \varepsilon_g=
\frac{\varepsilon_0}{g_{r0}}, \label{r15}
\end{eqnarray}
where the value $g_{r0}$ is defined in Eq.(\ref{r4}), $u$ is defined
in Eq.(\ref{r1}). Using Eqs.(\ref{7}), (\ref{8}), (\ref{r4}) one has
\begin{equation}
\frac{dI^{cr}}{d\omega}=\frac{4Z^2\alpha^3}{m^2}n_aL_0d_r(y_r)g_{r0}=
L_{rad}^{-1}\left(g_{r0}-\frac{1}{18 L_0}\right)d_r(y_r).
\label{r16}
\end{equation}
If one omits in this expression the crystal summand in
$g_{r0}~(h(u_1^2/a^2))$, the intensity $dI^{cr}$ will be very close
to the Bethe-Maximon one (see also Eq.(\ref{17})). In this case the
coherent contribution is
\begin{equation}
\frac{dI^{coh}}{d\omega}=\frac{\alpha m^2 x_m\sqrt{3}}{\varepsilon_m
x_0} (1-y_r+y_r^2)e^{-2u/3\chi_m}, \label{r17}
\end{equation}
where $x_m$ is defined in Eq.(\ref{15}).

The next terms of decomposition of the radiation intensity
$dI=dI(\varepsilon, y_r)$ over $\nu_0^2$ which defines the LPM
effect (compare with Eq.(\ref{18})) is
\begin{equation}
\frac{dI^{(4)}}{d\omega}\simeq \frac{dI^{cr}}{d\omega}
\frac{\varepsilon^2}{3u^2 \varepsilon_g^2} \left(1-\frac{20}{63}
\frac{(1-y_r)}{d_r(y_r)}\right)\left(1+80
\frac{\chi_m^2}{u^2}\right). \label{r18}
\end{equation}
The last expression has rather narrow interval of applicability
because of large coefficient 80 in front of depending on field
correction. In Eq.(\ref{r17}) we used the simple estimate
$-4f''(x_m)/f(x_m) \simeq 1/x_m^2~(x_m \simeq \eta \ll 1)$ and in
Eqs.(\ref{r15}) and (\ref{r18}) we used Eq.(\ref{19}) without
violating accuracy of derived above approximate expressions.

The spectra of radiation from an electron in tungsten, axis $<111>$,
temperature T=293 K, for the energies where the coherent
$I^{coh}(\varepsilon)$ and the incoherent $I^{inc}(\varepsilon)$
contributions to the total intensity are comparable, are shown in
Fig.3(a). These spectra describe radiation in thin targets where one
can neglect the energy loss of projectile. Weak variation of the
spectral intensity of radiation near the maximum in the soft part of
spectrum is described quite satisfactory by Eqs.(\ref{r10}),
(\ref{r12}). Although it is quite difficult to determine the
position of the maximum within a good accuracy, its height is given
by Eq.(\ref{r12}) with precision better 10 \%. It is seen that the
phenomena under consideration become apparent at relatively low
energy. For $\varepsilon=1$~GeV, $dI^{coh} \simeq dI^{inc}$ at
$y_r=y_c \simeq 0.2~(\omega \simeq 200~$MeV) while for lower photon
energy the coherent contribution dominates and for higher photon
energy the incoherent contribution dominates. For
$\varepsilon=3$~GeV, $dI^{coh} \simeq dI^{inc}$ at $y_r=y_c \simeq
0.42~(\omega \simeq 1.26$~GeV), for $\varepsilon=5$~GeV, $dI^{coh}
\simeq dI^{inc}$ at $y_r=y_c \simeq 0.54~(\omega \simeq 2.7$~GeV),
and for $\varepsilon=10$~GeV, $dI^{coh} \simeq dI^{inc}$ at $y_r=y_c
\simeq 0.7~(\omega \simeq 7$~GeV). One can estimate the position
$y_c (u_c=y_c/(1+y_c))$ using Eq.(\ref{r17})
\begin{equation}
\frac{dI^{coh}}{d\omega}(u=u_c) \sim\frac{\alpha m^2
\eta_1\sqrt{3}}{\varepsilon_m x_0}e^{-2u_c/3\chi_m}=L_{rad}^{-1}
=\frac{\alpha m^2 \eta_1}{4\pi\varepsilon_0 x_0},\quad
u_c=\frac{3\varepsilon}{2\varepsilon_m}\ln\frac{4\pi \sqrt{3}
\varepsilon_0}{\varepsilon_m}. \label{r19}
\end{equation}
The values of $y_c$ calculated according Eq.(\ref{r19}) are in a
good agreement with Fig.3(a).

We define the contribution of the LPM effect into the radiation
spectrum by analogy with \cite{BK1}, as
\begin{equation}
\Delta_r(\varepsilon, y_r)=-\frac{dI(\varepsilon, y_r)
-dI^{coh}(\varepsilon, y_r)-dI^{inc}(\varepsilon,
y_r)}{dI(\varepsilon, y_r)}. \label{r8}
\end{equation}
The function $\Delta_r(\varepsilon, y_r)$ is shown in Fig.3(b). It
reaches the highest value $\Delta_r=$ 6.03 \% at
$\varepsilon=10~$GeV and $y_r=0.82$. At $\varepsilon=5~$GeV the
maximal value $\Delta_r=$ 5.84 \% is attained at $y_r=0.68$, at
$\varepsilon=3~$GeV the maximal value $\Delta_r=$ 5.67 \% is
attained at $y_r=0.56$, and at $\varepsilon=1~$GeV the maximal value
$\Delta_r=$ 5.41 \% is attained at $y_r=0.3$.

All the curves in Fig.3(b) have nearly the same height of the
maximum and the position of the maximum $y_m$ and its magnitude are
defined roughly by the expressions $u_m \simeq
6\varepsilon/\varepsilon_m (y_m=u_m/(1+u_m)),~\Delta_r^{max} \simeq
\varepsilon_m^2/(48\varepsilon_g^2)$.

\section{Conclusion}

In this paper the spectral distribution of electron-positron pair
created by photon and the spectral distribution of radiation from
high-energy electron moving in an oriented crystal is calculated for
intermediate energies (a few GeV for heavy elements and a few tens
GeV for germanium). The interplay of the coherent and the incoherent
parts of corresponding process is essential for the summary
spectrum. Just in this situation the effects of multiple scattering
of charged particles appear.

In an oriented crystal at motion of created particles (or the
initial electron) near a chain of atoms (an axis) the atom density
on the trajectory is much higher than in an amorphous medium.
Because of this, the parameter, characterizing the influence of
multiple scattering on the pair creation process in a medium in
absence of an external field ($\nu_0^2 = \omega/\omega_0$), becomes
of the order of unity at relatively low energy (values of $\omega_0$
for tungsten and germanium are given in Table 1). For the radiation
process the characteristic energy $\varepsilon_0=\omega_0/4$. From
the other side, due to the high density of atoms on the trajectory
of created particles near the axis, the strong electric field of the
axis acts on the electron (positron). As a result, with energy
increase the pair creation formation length diminishes and the
characteristic angles of the process expand. Hence the influence of
multiple scattering on the process decreases. So, one has to use the
general expression for the pair creation probability, which includes
both the crystal effective field (the coherent mechanism) and the
multiple scattering (the incoherent mechanism) to study the pair
creation process in oriented crystal \cite{BK0}. The corresponding
expression for the radiation process was obtained in \cite{BK1}.

The two first terms of decomposition of the spectral probability of
pair creation $dW(\omega, y)$ over the parameter $\nu_0^2$ give the
coherent and the incoherent pair creation probabilities. It should
be noted that in the incoherent contribution the influence of the
axis field is taken into account. The next terms of the
decomposition represent the multiple scattering effect (the LPM
effect) in the presence of crystalline field.

Since in an amorphous medium even for heavy elements the LPM effect
in pair creation process can be observed only in TeV energy range
(see e.g. \cite{BK}), the possibility to study this effect in GeV
energy range is evidently of the great interest. The same is true
for the hard part of the radiation spectrum.

In the present paper the detailed analysis of the spectral
properties of the pair creation and radiation processes is
performed. The influence of different mechanisms on the general
picture of event is elucidated. At high energy $\omega \gg
\omega_m~(\varepsilon \gg \varepsilon_m)$ the influence of the
multiple scattering on the process under consideration (the LPM
effect) manifests itself for relatively low energy of one of the
final charged particles ($\varepsilon_f \sim \varepsilon_m \ll
\omega~(\varepsilon)$). In this region of spectrum $s_3(r_3) \simeq
0,~s_2(r_2) \simeq 1$ (see Eqs.(\ref{3}), (\ref{r2})), so that one
can present Eqs.(\ref{2}) and (\ref{r1}) in the form
\begin{eqnarray}
&&\frac{dW}{dy}=s_2(y)R_2(\omega y(1-y))-s_3(y)R_3(\omega y(1-y))
\simeq R_2(\varepsilon)=R_2(\varepsilon_f),
\nonumber \\
&& \frac{dI}{d\omega}=
r_2(y_r)R_{2}\left(\frac{\varepsilon}{u}\right)
+r_3(y_r)R_{3}\left(\frac{\varepsilon}{u}\right) \simeq
R_{2}(\varepsilon-\omega)=R_2(\varepsilon_f),\label{c1}
\end{eqnarray}
where we neglect the very small difference of the interpolating
functions $g_r(x)$ and $g_p(x)$ ($\sim 1$ \%). So we have the
scaling (dependence on the fixed combination of kinematic variables)
not only for different energies of the initial particles in a given
process, but also in the both crossing processes under consideration
since this is the same combination $\omega y(1-y) =\varepsilon/u =
\varepsilon(\varepsilon-\omega)/\omega$. For this reason at high
energy of the initial particles the maximum value of the LPM effect
for both processes is defined by the maximum of the function
$\Delta_{max}=\Delta(z_m)$, where
\begin{equation}
\Delta(z)=\frac{R_2^{coh}(z)+R_2^{inc}(z)}{R_2(z)}-1,\quad z_m
\simeq \frac{\varepsilon_m}{6}. \label{c2}
\end{equation}
In the low energy region $\omega(\varepsilon) \leq
\omega_m=\varepsilon_m$ this scaling remains only approximate one.
Nevertheless the value of maximum and its position vary weakly. Just
this energy region is suitable for the experimental study because
the rather wide of spectrum $\Delta y \sim 1$ contributes. It should
be emphasized that the LPM effect is large enough for heavy elements
only (it is around 8 \% in the maximum for tungsten at T=100 K, see
Fig.1(b)).

\vspace{0.5cm}

{\bf Acknowledgments}

The authors are indebted to the Russian Foundation for Basic
Research supported in part this research by Grant 06-02-16226.

\newpage

{\bf Figure captions}

{\bf Fig.1}~ The spectral distribution of created by a photon pair
vs the electron energy $y=\varepsilon/\omega$ in
tungsten, axis $<111>$, temperature T=100 K.\\
(a) The different contributions to the electron (positron) spectrum
(in units ${\rm cm}^{-1}$)  The curves 1, 2, 3, 4 are the theory
prediction  $dW(\omega, y)/dy$ (see Eq.(\ref{2})) for photon
energies $\omega=5,~7,~10,~15$~GeV respectively, The doted curves
1c, 2c, 3c, 4c are the corresponding coherent contributions
$dW^{coh}(\omega, y)/dy$, the dashed curves present the incoherent
contributions $dW^{inc}(\omega, y)/dy$. At $y \rightarrow 0.5$ these
curves from top to bottom are correspondingly for the photon
energies
$\omega=5,~7,~10,~15$~GeV.\\
(b) The relative contribution of the LPM effect in the spectral
distribution of created electron (see Eq.(\ref{13}))
$\Delta_p(\omega, y)$ (per cent). The curves 1, 2, 3, 4 are
correspondingly for photon energies $\omega=5,~7,~10,~15$~GeV.

{\bf Fig.2}~The spectral distribution of created by a photon pair vs
the electron energy $y=\varepsilon/\omega$ in
germanium, axis $<110>$, temperature T=293 K.\\
(a) The different contributions to the electron (positron) spectrum
(in units ${\rm cm}^{-1}$)  The curves 1, 2, 3,  are the theory
prediction  $dW(\omega, y)/dy$ (see Eq.(\ref{2})) for photon
energies $\omega=55,~75,~95$~GeV respectively, The doted curves 1c,
2c, 3c are the corresponding coherent contributions
$dW^{coh}(\omega, y)/dy$, the dashed curves present the incoherent
contributions $dW^{inc}(\omega, y)/dy$.  At $y \rightarrow 0.5$
these curves from top to bottom are correspondingly for the photon
energies
$\omega=55,~75,~95$~GeV.\\
(b) The relative contribution of the LPM effect in the spectral
distribution of created electron (see Eq.(\ref{13}))
$\Delta_p(\omega, y)$ (per cent). The curves 1, 2, 3 are
correspondingly for photon energies $\omega=55,~75,~95$~GeV.

{\bf Fig.3}~  The radiation spectral intensity vs the photon energy
$y= \omega/\varepsilon$ in tungsten, axis $<111>$,
temperature T=293 K.\\
(a) The intensity distribution  $dI(\varepsilon, y_r)/d\omega$ (in
units ${\rm cm}^{-1}$) The curves 1, 2, 3, 4 are the theory
prediction (see Eq.(\ref{r1})) for electron energies
$\varepsilon=1,~3,~5,~10$~GeV
respectively.\\
(b) The relative contribution of the LPM effect in the spectral
distribution of emitted photons (see Eq.(\ref{r8})) $\Delta_r$ (per
cent). The curves 1, 2, 3, 4 are correspondingly for the electron
energies $\varepsilon=1,~3,~5,~10$~GeV.

\newpage
\begin{table}
\begin{center}
{\sc Table 1}~ {Parameters of the pair photoproduction and radiation
processes in the tungsten crystal, axis $<111>$ and the germanium
crystal, axis $<110>$ for two temperatures T
($\varepsilon_0=\omega_0/4, \varepsilon_m=\omega_m,
\varepsilon_s=\omega_s$)}
\end{center}
\begin{center}
\begin{tabular}{*{10}{|c}|}
\hline Crystal& T(K)&$V_0$(eV)&$x_0$&$\eta_1$&$\eta$&
$\omega_0$(GeV)&$\varepsilon_m$(GeV)&$\varepsilon_s$(GeV)&$h$ \\
\hline W & 293&417&39.7&0.108&0.115&29.7&14.35&34.8&0.348\\
\hline W &100&355&35.7&0.0401&0.0313&12.25&8.10&43.1&0.612\\
\hline Ge & 293 & 110& 15.5
&0.125&0.119&592&88.4&210&0.235\\
\hline Ge & 100 & 114.5& 19.8
&0.064&0.0633&236&50.5&179&0.459\\
\hline
\end{tabular}
\end{center}
\end{table}


\newpage

\begin{thebibliography}{99}
\bibitem{ABB} P. L. Anthony, R. Becker-Szendy, P. E. Bosted {\em et al},
Phys.Rev.{\bf D 56} (1997) 1373.
\bibitem{HU} H. D. Hansen, U. I. Uggerhoj, C.C.Biino {\em et al},
Phys.Rev. {\bf D 69} (2004) 032001.
\bibitem{BK} V. N. Baier and V. M. Katkov,
Phys.Rep. {\bf 409} (2005) 261.
\bibitem{BK0} V. N. Baier, and V. M. Katkov,
Phys.Lett., {\bf A 346} (2005) 359.
\bibitem{BK1} V. N. Baier, and  V. M. Katkov,
Phys. Lett.,A {\bf 353} (2006) 91.
\bibitem{NA63} J. U. Andersen, K.Kirsebom, S. P. Moller {\em et al},
{\em Electromagnetic Processes in Strong Cristalline Fields},
CERN-SPSC-2005-030.
\bibitem{BKS} V. N. Baier, V. M. Katkov and V. M. Strakhovenko,
{\em Electromagnetic Processes at High Energies in Oriented Single
Crystals} (World Scientific Publishing Co, Singapore, 1998).



\end{thebibliography}
\end{document}